\documentclass[a4paper,twocolumn,11pt,accepted=2026-03-30]{quantumarticle}
\pdfoutput=1
\usepackage[utf8]{inputenc}
\usepackage[english]{babel}
\usepackage[T1]{fontenc}
\usepackage{amsmath}
\usepackage{amssymb}
\usepackage{braket}
\usepackage{hyperref}
\usepackage{nccmath}
\usepackage{float}

\begin{document}

\title{Fibonacci Waveguide Quantum Electrodynamics}

\author{Florian B\"onsel}
\affiliation{Max Planck Institute for the Science of Light, 91058 Erlangen, Germany}
\affiliation{Department of Physics, Friedrich-Alexander-Universit\"at Erlangen-N\"urnberg, 91058 Erlangen, Germany}
\email{florian.boensel@mpl.mpg.de}

\author{Flore K.~Kunst}
\affiliation{Max Planck Institute for the Science of Light, 91058 Erlangen, Germany}
\affiliation{Department of Physics, Friedrich-Alexander-Universit\"at Erlangen-N\"urnberg, 91058 Erlangen, Germany}

\author{Federico Roccati}
\affiliation{Max Planck Institute for the Science of Light, 91058 Erlangen, Germany}
\affiliation{Quantum Theory Group, Dipartimento di Fisica e Chimica -- Emilio Segr\`e, Universit\`a degli Studi di Palermo, via Archirafi 36, I-90123 Palermo, Italy}

\maketitle

\begin{abstract}
  Waveguide quantum electrodynamics (QED) provides a powerful framework for engineering quantum interactions, traditionally relying on periodic photonic arrays with continuous energy bands. Here, we investigate waveguide QED in a fundamentally different environment: A one-dimensional photonic array whose hopping strengths are structured aperiodically according to the deterministic Fibonacci-Lucas substitution rule. These ``Fibonacci waveguides'' lack translational invariance and are characterized by a singular continuous energy spectrum and critical eigenstates, representing a deterministic intermediate between ordered and disordered systems. We demonstrate how to achieve decoherence-free, coherent interactions in this unique setting. We analyze two paradigmatic cases: (i) Giant emitters resonantly coupled to the simplest aperiodic version of a standard waveguide. For these, we show that atom-photon bound states form only for specific coupling configurations dictated by the aperiodic sequence, leading to an effective atomic Hamiltonian, which itself inherits the Fibonacci structure; and (ii)  emitters locally and off-resonantly coupled to the aperiodic version of the Su-Schrieffer-Heeger waveguide. In this case the mediating bound states feature aperiodically modulated profiles, resulting in an effective Hamiltonian with multifractal properties. Our work establishes Fibonacci waveguides as a versatile platform, which is experimentally feasible, demonstrating that the deterministic complexity of aperiodic structures can be directly engineered into the interactions between quantum emitters.
\end{abstract}

	\begin{figure*}
		\centering
\includegraphics[width=\textwidth]{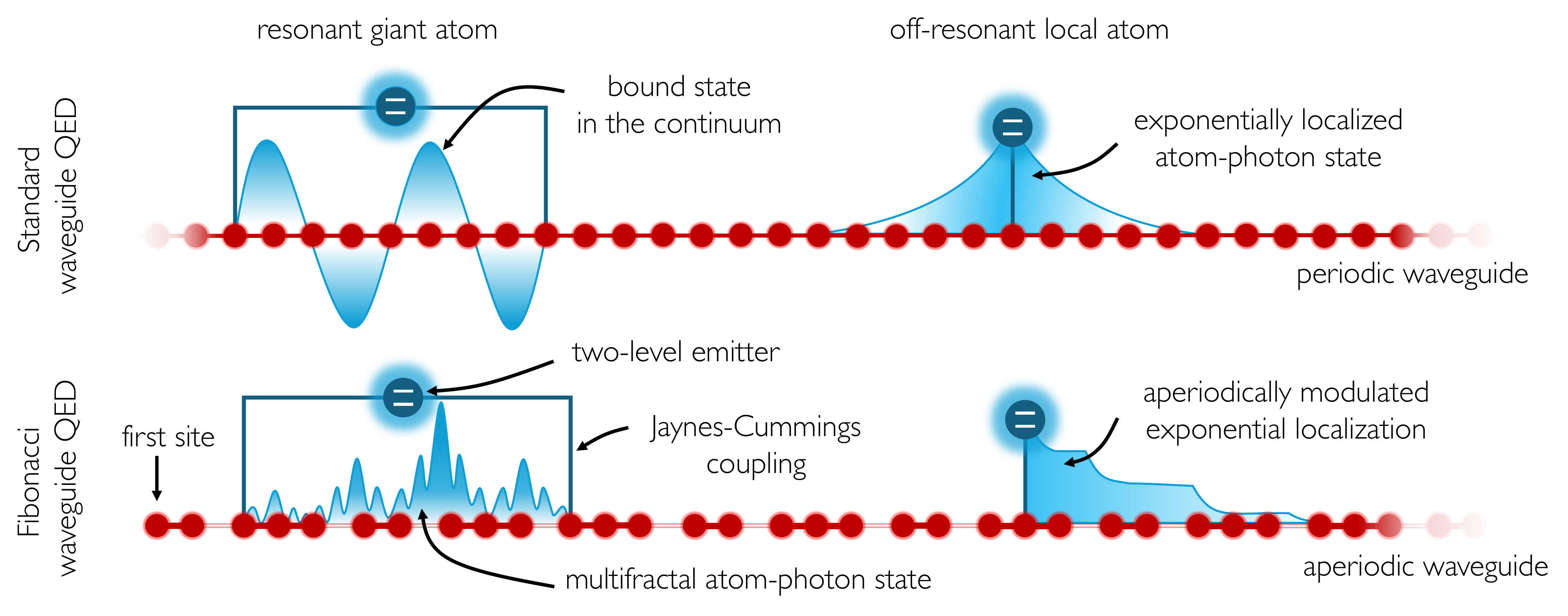}
		\caption{ \textit{Standard vs Fibonacci waveguide QED}.
        Quantum emitters [blue] coupled to one-dimensional arrays of resonators (waveguides) [red]. 
        The photon-mediated decoherence-free interactions between emitters arise from virtual photons that are strictly localized between the coupling points of the giant (multilocal) emitters in the resonant case [left], or from localized virtual photonic clouds in the vicinity of off-resonant locally-coupled atoms [right].
        In standard waveguide QED (top row), for which the sequence of hopping rates is periodic,
        the former are standing waves between the coupling points (bound states in the continuum, BICs), and the latter are exponentially localized states (near the emitter's position) .
        In Fibonacci waveguide QED (bottom row), for which the sequence of hopping rates is aperiodic (as indicated by the thick and light red bonds), the BICs turn into multifractal photonic states. The (local) atom-photon bound states are chiral (as in the SSH waveguide) and their localization profiles are aperiodically modulated by the underlying waveguide hopping pattern. The figure is intended to convey qualitative insights rather than quantitative results.
        }
		\label{fig:illustration}
	\end{figure*}

    Waveguide quantum electrodynamics (QED) has emerged as a cornerstone for modern quantum technologies, enabling the control of light-matter interactions 
    \cite{ yalla2014cavity, goban2014atom,albrecht2019subradiant, perczel2017topological,   sheremet2023waveguide, ciccarello2024waveguide}. 
    By coupling quantum emitters (two-level artificial atoms) to one-dimensional photonic reservoirs, it is possible to mediate long-range interactions \cite{wang2024long, garcia2020tunable}, 
    generate entanglement \cite{mirza2016multiqubit, yin2023generation}, and study collective quantum phenomena such as super- and subradiance \cite{dicke1954coherence, manzoni2018optimization, cardenas2023many, albrecht2019subradiant}. 
    The vast majority of theoretical and experimental work in this field has focused on waveguides with periodic structures, such as uniform tight-binding arrays or the Su-Schrieffer-Heeger (SSH) model \cite{ryu2002topological, asboth2016short}. 
    The translational invariance of these systems leads to well-defined energy bands of extended Bloch waves.
    By coupling quantum emitters to such arrays, photon-mediated interactions emerge between them. These interactions are either dissipative and long-range when the emitters are resonant with a band, or coherent and short-range when they are resonant with a bandgap \cite{bello2019unconventional,sanchez2020limits}.

    Departing from perfect periodicity opens a new frontier for controlling quantum interactions \cite{macia2005role,lapierre2025entanglement}. While disorder-induced Anderson localization \cite{anderson1958absence} has been studied extensively \cite{evers2008anderson, segev2013anderson, chen2024anderson, jiricek2025universal}, the intermediate regime of quasiperiodicity offers a unique, deterministically complex environment. 
    The Aubry-Andr\'e-Harper (AAH) model~\cite{aubry1980analyticity, harper1955single, gordon1997duality, kraus2012topological, hetenyi2025numerical} with its quasiperiodic onsite potential is a canonical example famous for its sharp localization transition~\cite{kohmoto1983metal, sarma1990localization}. Similarly, quasiperiodicity can be engineered into the hopping terms of a Hamiltonian, yielding a system which is also neither translationally invariant nor truly disordered~\cite{jagannathan2021fibonacci}.

    In this work, we introduce and explore a novel platform for waveguide QED based on such photonic arrays, which we dub \textit{Fibonacci waveguides}. The Hamiltonian of this waveguide is a tight-binding model in which the nearest-neighbor hopping rates are arranged according to the aperiodic, though deterministic, Fibonacci-Lucas substitution rule \cite{lucas1875, ballot2023lucas,jagannathan2021fibonacci}. Unlike their periodic counterparts, these waveguides feature a singular continuous energy spectrum with a multifractal structure and eigenstates which are critical -- neither fully extended nor exponentially localized \cite{jagannathan2021fibonacci, piechon1995analytical, macia1996physical, jianxin1993green, fujiwara1989multifractal, mace2016fractal, kohmoto1983localization, ostlund1983one, reisner2023experimental,schmid2024self,kobialka2024topological,tanese2014fractal, moustaj2023spectral,sandberg2024josephson, PoddubnyPRB2008, HendricksonOE2008, PoddubnyPRB2009, qi2023linear}. We will use the Fibonacci sequence as the aperiodic analogue of the uniform tight-binding waveguide model, and the first nontrivial Lucas sequence as the aperiodic version of the SSH waveguide.
    
We demonstrate that these Fibonacci waveguides can be harnessed to engineer unique decoherence-free quantum interactions mediated by atom-photon bound states~\cite{bykov1975spontaneous, john1990quantum, kurizki1990two, karg2019remote, leonforte2024quantum}. We investigate two distinct physical scenarios. First, we couple ``giant''  atoms, which are coupled to the waveguide at multiple points~\cite{takagi2021international, wang2022giant, gonzalez2019engineering, kockum2021quantum, wang2024realizing, wang2024nonlinear}, to the Fibonacci waveguide based on the Fibonacci sequence. We show that the formation of bound vacancy-like dressed states (VDSs) \cite{leonforte2021vacancy} is critically dependent on both the distance between coupling points and their precise location within the aperiodic sequence. This conditional binding gives rise to an effective Hamiltonian for the emitters, which inherits the aperiodic Fibonacci structure of the underlying waveguide while also featuring long-range interactions. Second, we couple standard ``small'' atoms to the gapped Fibonacci waveguide based on the first nontrivial Lucas sequence. The resulting atom-photon bound states exhibit spatial profiles, which are aperiodically modulated, leading to effective multifractal dipole-dipole interactions, see Fig.~\ref{fig:illustration}.

We show that these coherent interactions can be accurately described by the physics of atom-photon bound states, even though the nature of the waveguide's density of states makes the derivation of a standard master equation intractable (remarkably for any value of the atom-field detuning). Our results
establish quasiperiodic systems as a versatile platform for mediating complex interactions with distinctive length scales.

\section{The Fibonacci Waveguide}\label{sec:fibwg}
	
	With the present study we investigate a waveguide QED system made of quantum emitters coupled to 
	a (class of)
	one-dimensional~(1D) aperiodic photonic arrays, which we dub \textit{Fibonacci-Lucas}
    (or shortly \textit{Fibonacci}) \textit{waveguides}. In this section we illustrate the main properties of the 1D photonic array, whose Hamiltonian reads  
	\begin{equation}\label{eq:waveguide}
		\hat H_{(p,q)} 
		= 
		\omega_c
		\sum_{n} 
		\hat a_n^{\dagger}
		\hat a_n 
		+ 
		\sum_{n}
		t_n
		\left( 
		\hat a_{n+1}^{\dagger}
		\hat a_{n} 
		+ 
		{\rm H.c.}
		\right).
	\end{equation}
	This Hamiltonian describes a set of $N$ resonators (bosonic modes) $\hat a_n$, $n=0,\ldots,N-1$, with bare frequency $\omega_c$ and position-dependent nearest-neighbor hopping rates $t_n$. 
	The model is aperiodic as the values of the  sequence $\{ t_0, t_1,t_2,\ldots \}$ are chosen following the $(p,q)$-Lucas substitution rule, which we now describe.
	
	The $(p,q)$-Lucas substitution rule is a generalization of the Fibonacci substitution rule (obtained for ${p=q=1}$) used to generate aperiodic sequences  of two symbols \cite{lucas1875,jagannathan2021fibonacci}.
	Considering the two symbols $A$ and $B$, the  Lucas substitution rule $\sigma$ is given by
	\begin{equation}\label{eq:rule}
		\sigma:
		\begin{cases}
			A\,\,\,  \mapsto \,\,\,  A^pB\\
			B \,\,\, \mapsto \,\,\, A^q
		\end{cases},
	\end{equation}
	with $p$ and $q$ positive integers, and ${A^p=AA...A}$ ($p$ times). We refer to a (sub-)sequence of symbols as (sub-)word. As an example, in the Fibonacci case ($p=q=1$), we get ${\sigma(A) = AB}$, ${\sigma^2(A) = \sigma(\sigma(A)) = ABA}$, ${\sigma^3(A) = ABAAB}$, and so on. The length of these words $\{ 2, 3, 5, ... \}$ correspond to the Fibonacci numbers (Lucas numbers for arbitrary $p$ and $q$~\cite{lucas1875, ballot2023lucas}). 
	Identifying $(A,B)$ with a pair of numbers $(t_A,t_B)$, one can construct an aperiodic sequence of hoppings, which follows the $(p,q)$-Lucas rule. As an example, the Fibonacci rule gives the sequence of hoppings $\{ t_A,t_B,t_A,t_A,t_B,\ldots \}$. 
	Eventually, the relevant parameter in the  waveguide Hamiltonian is $\rho = t_B/t_A$.
	In this manuscript, we study the case of a Fibonacci waveguide $\hat H_{(1,1)} $ corresponding to the Fibonacci sequence, and the first non-trivial waveguide $\hat H_{(1,2)} $ (the reason will be clear shortly). 
	
	It will be useful to compare our model to other waveguide  models, in particular the uniform tight-binding and the Su-Schrieffer-Heeger (SSH) model. The former is obtained by Eq.~\eqref{eq:waveguide} from setting $t_n=t_A$ for all $n\geq 0$ (or equivalently, $\rho=1$), the latter from setting $t_{2n}= t_A$ and $t_{2n+1}= t_B$.
	The uniform model, whose Hamiltonian we label $\hat H_{\rm uni}$, features a dispersive band of width $4t_A$ centered around $\omega_c$ which supports extended Bloch waves. The SSH model, $\hat H_{\rm SSH}$, differs (in the thermodynamic limit) from the uniform model in that a bandgap of width $2|t_A - t_B|$ opens up around $\omega_c$~\cite{LombardoPRA2014,bello2019unconventional}.
	In a heuristic sense and with a focus on the presence of a gap around $\omega_c$, the Fibonacci waveguide $\hat H_{(1,1)} $ and $\hat H_{(1,2)}$ are the aperiodic versions of the uniform and SSH models, respectively, see Fig.~\ref{fig:dos} (and Appendix~\ref{app:SSH1112} for more details).

	A third model we will use for benchmarking is the Aubry-Andr\'e-Harper (AAH) model, whose Hamiltonian reads
	\begin{equation}
		\hat H_{\rm AAH}=
		\!
		\sum_{n}
		2V\!
		\cos(2\pi\beta n) 
		\hat a_n^{\dagger}
		\hat a_n 
		\!+ 
		t
		( 
		\hat a_{n+1}^{\dagger}
		\hat a_{n} 
		\!+ 
		\!{\rm H.c.}
		).
		\label{eq:AAH}
	\end{equation}
	with $\beta$ an irrational number. This model features three regimes: A delocalized ($V/t<1$) and a localized  ($V/t>1$) phase, separated by a point ($V/t=1$) at which the spectrum and the eigenmodes are multifractal. The latter are key features of our Fibonacci waveguide.
	
	\subsection{General remarks}
	
	Before describing the salient features of the  Fibonacci waveguide, we make 
    some general remarks to motivate our study and place it in the context of current waveguide QED literature.
	
	\begin{itemize}
		\item For \textit{periodic} sequences of hoppings one can take the thermodynamic limit $N\to\infty$ (or consider the waveguide under periodic boundary conditions) and exploit translational invariance by using the Bloch theorem to diagonalize the Hamiltonian \cite{joannopoulos2008molding}. 
		In contrast, a Fibonacci waveguide features a first site ($n = 0$), lacks translational invariance, and has effectively open boundaries. Taking the thermodynamic limit for a Fibonacci waveguide amounts to generating larger and larger aperiodic sequences of hoppings using repeatedly the Lucas substitution rule \eqref{eq:rule}.
		\item We recall that any tight-binding lattice Hamiltonian featuring solely nearest-neighbor hoppings enjoys chiral symmetry, due to its bipartite nature \cite{ryu2002topological}.
		Therefore, the Fibonacci waveguide Hamiltonian (up to the onsite energy term) has chiral symmetry. This implies that the spectrum is symmetric with respect to $\omega_c$.
		\item Photonic arrays routinely considered in waveguide QED feature a continuous spectrum made of bands and support delocalized Bloch eigenmodes with velocity \cite{joannopoulos2008molding}.
        These features, along with the assumption of weak atom-light coupling, justify referring to periodic photonic arrays as \textit{waveguides}.
		Therefore, the term \textit{waveguide} for aperiodic models needs some justification.
		First, note that there is a plethora of aperiodic 1D tight-binding models on top of the ones of Eq.~\eqref{eq:waveguide}, for instance the aforementioned AAH model in Eq.~\eqref{eq:AAH}:
		for $V/t<1$ the spectrum is continuous and the eigenstates are delocalized, while for $V/t>1$ the model features a point spectrum and the eigenstates are localized \cite{jagannathan2021fibonacci}. In the localized phase, transport is suppressed. Therefore it would be improper to refer to the AAH model in this phase as a waveguide. At the localization transition point $V/t=1$ the spectrum is \textit{multifractal} (more details below) and the eigenstates are \textit{critical} (neither extended nor localized).
		In stark contrast, our model in Eq.~\eqref{eq:waveguide} features a multifractal spectrum and critical (importantly, \textit{not} localized) eigenmodes for \textit{any} value of the parameters. Therefore, with a slight abuse of nomenclature, we can refer to Eq.~\eqref{eq:waveguide} as a Fibonacci \textit{waveguide}.
		\item The setup we propose is realizable with state-of-the-art technology. Similar waveguide QED setups, with photonic arrays featuring only nearest-neighbor couplings, have been recently realized with superconducting resonators~\cite{ferreira2021collapse, jouanny2025high, kim2021quantum, zhang2023superconducting, scigliuzzo2022controlling, owens2022chiral, carusotto2020photonic}.
	\end{itemize}

	In the reminder of this section we describe the main features of the single-particle spectrum and eigenmodes of the Fibonacci waveguide. 
	To the best of our knowledge, the properties of quasiperiodic models based on the Fibonacci sequence are well understood, while the same cannot be said for models based on the Lucas substitution rule.

	\subsection{Multifractality}
	In periodic tight-binding arrays under periodic boundary conditions, the single-particle spectrum can be analytically derived. Translational invariance yields a conserved quantity, the quasi-momentum $k$, which labels the modes of the system. For a uniform finite lattice with $N$ sites, the mode spacing is $\Delta k = 2\pi/N$,
	so that the spectrum is discrete.
	By taking the thermodynamic limit $N\to\infty$, the modes' spacing vanishes and the discrete energies form continuous bands.
	The single-particle eigenmodes can be labeled by the quasimomentum $k$ and the wavefunctions are simply plane waves $\propto e^{ikn}$.

    In sharp contrast, a Fibonacci waveguide, as many other quasiperiodic models, exhibits spectral properties which differ dramatically from both periodic and disordered systems \cite{jagannathan2021fibonacci, mace2017critical}. In the thermodynamic limit, its energy spectrum is neither continuous (as in periodic systems) nor pure point (as, e.g., in the localized AAH model), but rather \textit{singular continuous}, with a \textit{multifractal} structure.
	
	A multifractal spectrum is characterized by local power-law singularities in the density of states (DOS). More precisely, the integrated DOS, defined as $\mathcal N (E)$, counting the fraction of states with energy less than $E$, scales locally as $\mathcal N (E+\Delta E)- \mathcal N (E) \sim \Delta E ^{\alpha(E)}$. 
	Here, $\alpha(E)$ is a local H\"older exponent varying across the spectrum~\cite{jagannathan2021fibonacci}. Unlike for periodic systems, increasing the system size of the Fibonacci waveguide does not lead to smooth continuous bands; instead, the spectrum becomes increasingly fragmented: each energy cluster (or ``bandwidth'') shrinks in size, and different clusters scale with different exponents $\alpha$.
	In simpler terms, the multifractal nature of the spectrum reveals itself through a hierarchy of scaling behaviors: different parts of the spectrum respond differently under magnification, unlike a regular fractal, which scales uniformly~\cite{jagannathan2021fibonacci}, cf. Fig.~\ref{fig:dos}. 
	In Appendix~\ref{app:cantor} we provide more details on the multifractality of the models we consider.

    \begin{figure}[H]
		\includegraphics[width=\columnwidth]{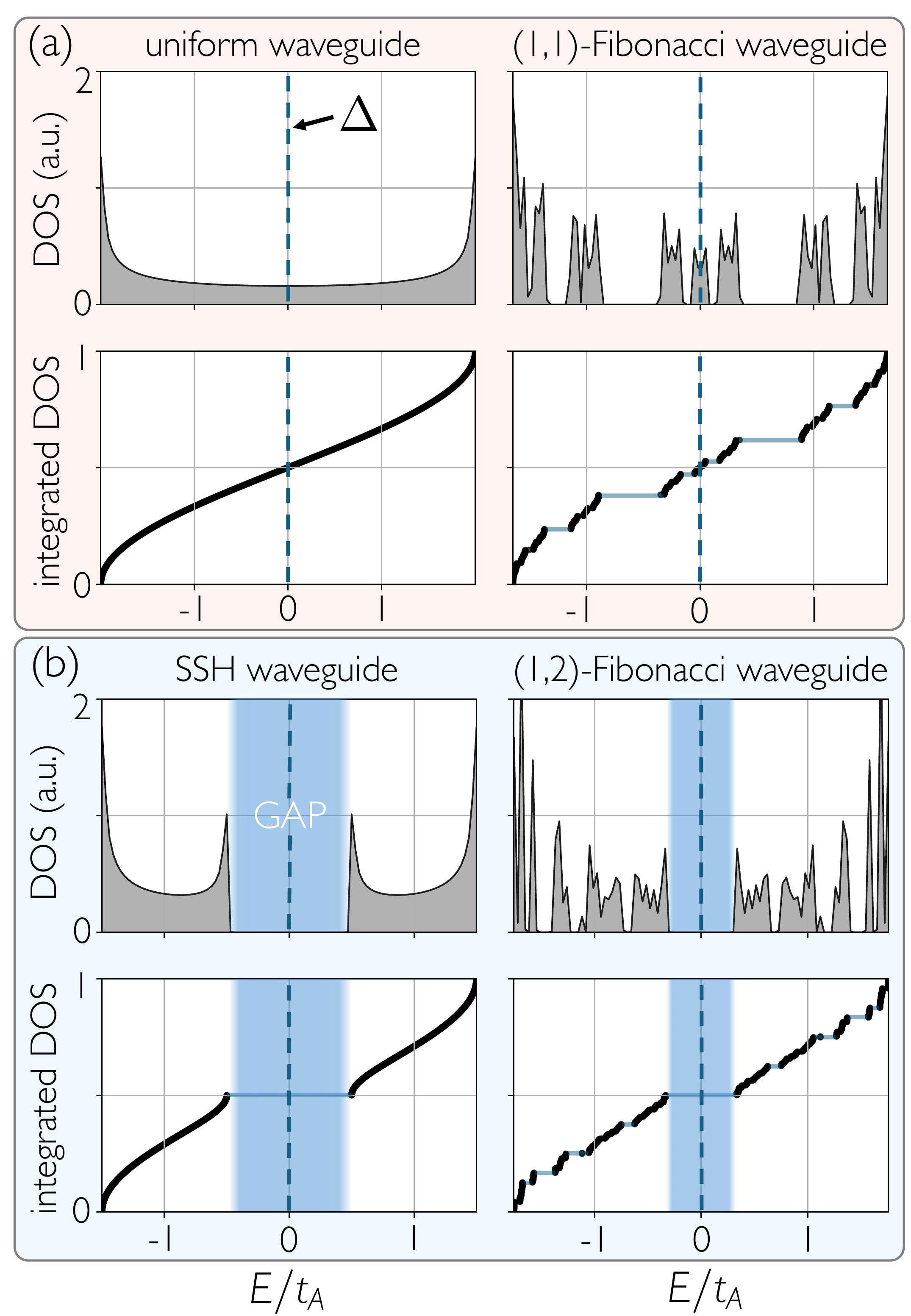}
		\caption{ \textit{Standard vs Fibonacci waveguides}. 
        In both panels (a) and (b) we show:  the density of states, DOS, (top row), and integrated DOS, $\mathcal N(E)$. Panel (a): uniform waveguide (left) and its aperiodic version, the $(1,1)$-Fibonacci waveguide (right). 
        Panel (b): SSH waveguide (left) and its aperiodic version, the $(1,2)$-Fibonacci waveguide (right). Notice that a central gap persist (shaded blue) in addition to all the fractal gaps~\cite{jagannathan2021fibonacci}.
        The vertical dashed line in all plots represents the atomic detuning $\Delta$, cf.~Eq.~\eqref{eq:FibWQEDHam}, which is set to zero throughout the manuscript. Parameter values: uniform waveguide, $t_B=t_A$; $(1,1)$-Fibonacci, $(1,2)$-Fibonacci, and SSH waveguides, $t_B=0.5t_A$.}
		\label{fig:dos}
	\end{figure}
	
\begin{figure}[h]
		\centering
		\includegraphics[width=\columnwidth]{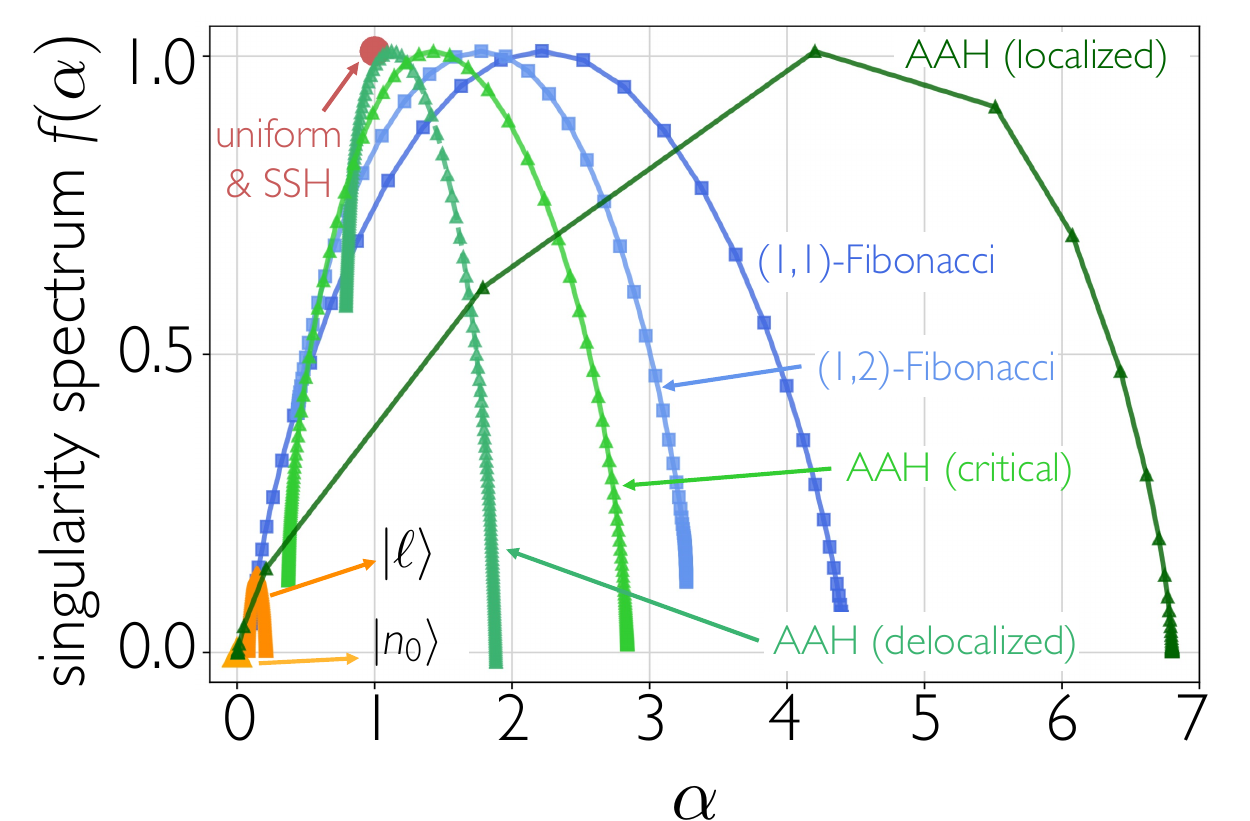}
		\caption{ \textit{Singularity spectrum}. 
        For delocalized (localized) models $f(\alpha)$ shrinks to a single point $f(1)=1$ shown in red ($f(0)=0$, yellow triangle). 
        Multifractal models have a continuum of scaling exponents.
        Parameter values: AAH model, $V=1.5\,t, t, 0.5\,t$ for the localized, critical, and localized phase, respectively; uniform waveguide, $t_B=t_A$; $(1,1)$-Fibonacci, $(1,2)$-Fibonacci, and SSH waveguides, $t_B=0.2t_A$. All models have a system size of $N=2048$, except the $(1,1)$-Fibonacci with $N=1597$ (a Fibonacci number).
        For each model we use a representative energy eigenstate $\ket{E_i}$, with $i$ denoting the index in the energy-ordered list of eigenstates. For the uniform and SSH waveguides under periodic boundary conditions, we use a band mode, namely $\ket{E_{1024}}_\text{uni}$ and $\ket{E_{1024}}_\text{SSH}$.
        For the delocalized, critical, and localized AAH model we use $\ket{E_{1220,\,1220,\,1220}}_\text{AAH}$, respectively.
        For the $(1,1)$- and $(1,2)$-Fibonacci waveguides the states $\ket{E_{798}}_\text{$(1,1)$}$ and $\ket{E_{1124}}_\text{$(1,2)$}$ are used, respectively. The orange curve shows the singularity spectrum of a normalized state $\ket{\ell}$, which is strictly localized (with same amplitude and phase) only on three neighboring sites. Lastly, $\ket{n_0}=\hat a^\dagger_{n_0}\ket{\text{vac}}$. Note that for the localized AAH model, $f(\alpha)$ takes a lot of data points around $\alpha=0$ as expected. There are also data points for large $\alpha$, which however, are a direct result of the exponentially decreasing tails of the localized state, see Appendix~\ref{app:falpha}.}
		\label{fig:singulspecIPR}
	\end{figure}

	A standard tool to measure multifractality is the singularity spectrum 
	$f(\alpha)$ \cite{fujiwara1989multifractal, kohmoto1987critical, holzer1991multifractal, mace2017critical, halsey1986fractal}, which gives the Hausdorff dimension of the set of points with local scaling exponent equal to 
	$\alpha$. 
	In physical terms, $f(\alpha)$ describes how densely the spectrum fills the energy axis at different scaling rates. For uniform systems such as the standard tight-binding model $\hat H_{\rm uni}$, the singularity spectrum \( f(\alpha) \) is trivial: the spectrum is continuous, and the integrated DOS scales uniformly with exponent \( \alpha = 1 \).
    This corresponds to a single point in the \( f(\alpha) \) curve, namely \( f(1) = 1 \), reflecting the homogeneous nature of the extended states and the regular density of energy levels.
	For localized models, such as the AAH model for $V/t>1$ (pure point spectrum and localized eigenstates), the integrated DOS exhibits step-like behavior, corresponding again to a trivial singularity spectrum. In this case, the integrated DOS has zero derivative almost everywhere, and the scaling exponent \( \alpha \to 0 \), indicating that the spectrum does not fill any region densely. The \( f(\alpha) \) curve in this case collapses to a single point at \( f(0) = 0 \) (this is actually achieved for a perfectly localized state $\ket{n_0}=\hat a^\dagger_{n_0}\ket{\text{vac}}$, cf.~Fig.~\ref{fig:singulspecIPR}).
    Owing to the multifractal nature of the spectrum of aperiodic models, there are no regions in the energy axis which fill densely in the thermodynamic limit. 
    Therefore, the singularity spectrum becomes a continuous curve in the multifractal case. In Fig.~\ref{fig:singulspecIPR}, we show the singularity spectrum of the uniform, SSH, and Fibonacci waveguides, as well as of the AAH model as a benchmark.
    In Appendix~\ref{app:falpha}, we describe how we computed the singularity spectrum.

	Remarkably, even though lacking translational invariance, certain Fibonacci waveguides support a (non-fractal) photonic bandgap, see Appendix~\ref{app:SSH1112} and~\ref{app:centralgap}. Indeed, when $p$ is odd and $q$ is even, the Fibonacci waveguide spectrum  has a gap centered around $\omega_c$, see Fig.~\ref{fig:dos}(b) and Fig.~\ref{fig:gapsize} in Appendix~\ref{app:centralgap}. For all other values of $p$ and $q$, the systems has no gap around $\omega_c$. For this reason (and others discussed later), we refer to the Fibonacci waveguide $\hat H_{(1,1)} $ ($\hat H_{(1,2)}$) as the aperiodic version of the uniform (SSH) model.
	
\begin{figure}[h]
		\centering
		\includegraphics[width=\columnwidth]{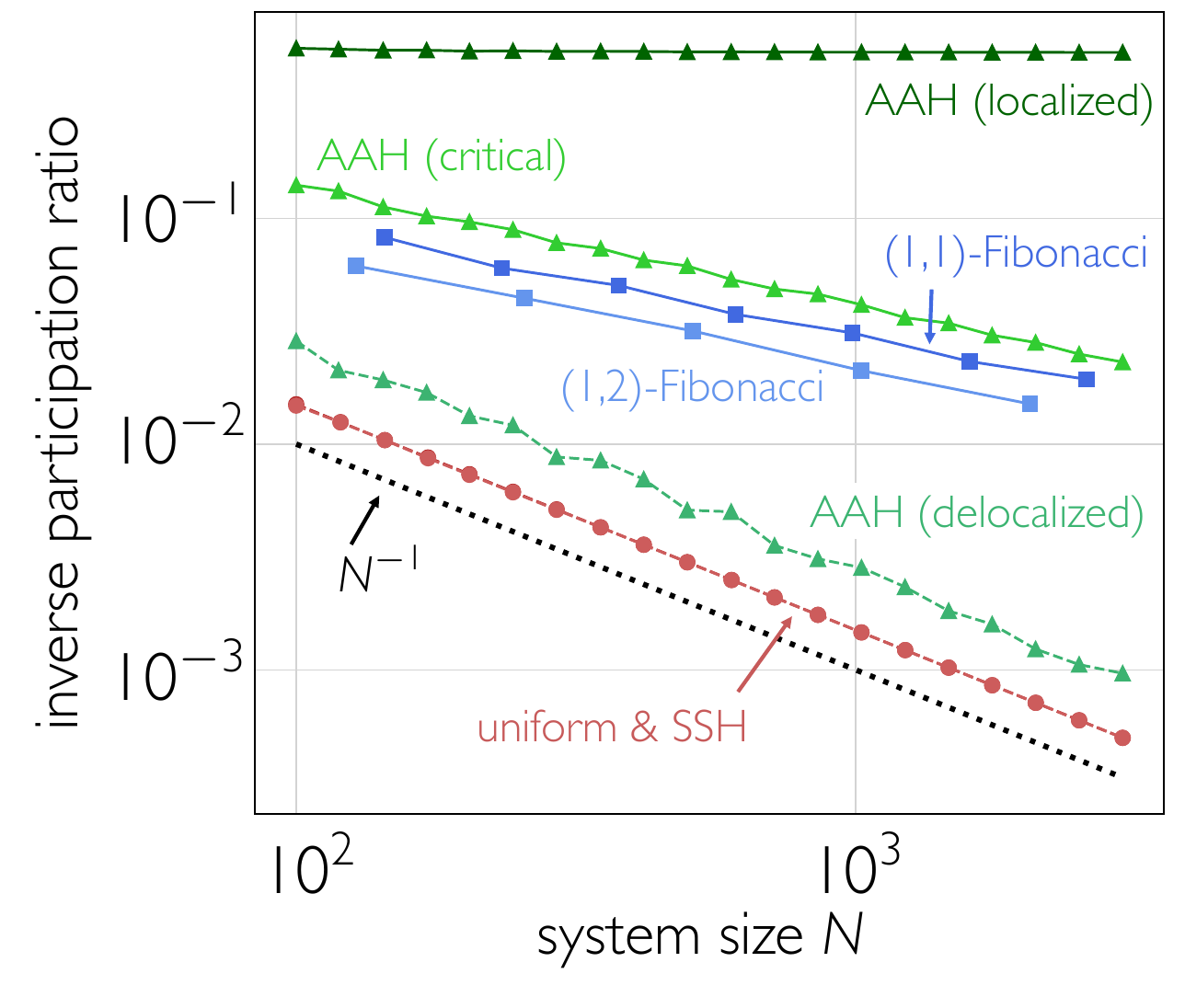}
		\caption{ \textit{Inverse Participation Ratio}. Inverse participation ratio averaged over all eigenstates versus system size. For delocalized models (as the uniform  and SSH waveguides under periodic boundary conditions, and the  AAH model with $V<t$) the IPR scales with system size $\propto N^{-1}$.
        For localized models (as the AAH model for $V>t$) the IPR is constant.
        Critical models (as the $(1,1)$- and $(1,2)$-Fibonacci waveguides, or the AAH model for $V=t$) display non-trivial scaling with system size. Parameter values: AAH model, $V=1.5\,t,t,0.5\,t$ for the localized, critical and localized phase, respectively; uniform waveguide $t_B=t_A$; $(1,1)$-Fibonacci, $(1,2)$-Fibonacci, and SSH waveguides, $t_B=0.2t_A$.}
		\label{fig:singulspecIPRR}
	\end{figure}

	A complementary, and possibly more immediate, signature of multifractality is the nontrivial scaling of the inverse participation ratio (IPR) with system size.
    The IPR is defined for a normalized eigenstate $\ket{E}$ as ${{\rm IPR}_E = \sum_n |\!\braket{n}{E}\!|^4}$ with $\ket{n}$ the site state, i.e., the single particle state $\hat a_n^\dagger\ket{{\rm vac}}$. In a system of size $N$, the IPR scales as $1/N$ (is constant) for extended (localized) states. 
    In Fig.~\ref{fig:singulspecIPRR}, we show the IPR of the uniform, SSH, and Fibonacci waveguides, as well as of the AAH model as a benchmark.

	\section{Quantum electrodynamics in a Fibonacci waveguide} \label{sec:qedfibwg}
	
	We now describe the full waveguide QED system. In a frame rotating at the bare resonators' frequency $\omega_c$, the full light-matter Hamiltonian reads 
	\begin{equation}\label{eq:FibWQEDHam}
		\hat H 
		=
		\Delta
		\sum_{j=1}^{N_e}
		\hat \sigma_j^\dagger \hat \sigma_j
		+ 
		\sum_{n}
		t_n
		\left( 
		\hat a_{n+1}^{\dagger}
		\hat a_{n} 
		+   
		{\rm H.c.}
		\right)
		+
		\hat H_{\rm int}\,.
	\end{equation}
	The first term describes a collection of $N_e$ artificial two-level atoms (quantum emitters).
	The operator $\hat \sigma_j = \ket{g}_j\!\bra{e}$ destroys an atomic excitation in atom $j$, $\ket{e}$ and $\ket{g}$ are the excited and ground states of the emitters, respectively, with transition frequency $\omega_e$, and $\Delta = \omega_e - \omega_c$ is the detuning of the atoms from the bare resonators' frequency.
	The second term in Eq.~\eqref{eq:FibWQEDHam} is the hopping part of the Fibonacci waveguide Hamiltonian in Eq.~\eqref{eq:waveguide} as the onsite energy term is included in the detuning. 
	The third term in Eq.~\eqref{eq:FibWQEDHam} is the interaction Hamiltonian $\hat H_{\rm int}$ in the rotating-wave approximation. 
	In this work we focus on decoherence-free interactions, which are mediated by Fibonacci waveguides, and we leave other questions, e.g., the (collective) radiance in aperiodic waveguides, for future studies. We consider two scenarios for the light-matter coupling.
	First, we consider the case of giant (i.e., multilocal) emitters coupled to $\hat H_{(1,1)}$ (corresponding to the Fibonacci sequence), so that the interaction Hamiltonian reads
	\begin{equation}\label{eq:HintGiant}
		\hat H_{\rm int}
		=
		g
		\sum_{j=1}^{N_e}
		\left[
		\hat \sigma_j
		\left(
		\hat a_{n_j} ^\dagger
		+
		\hat a_{n_j + d_j} ^\dagger
		\right)
		+
		{\rm H.c.}
		\right].
	\end{equation}
	Here we are considering only two-legged giant atoms, i.e., each atom $j$ is coupled to two resonators at positions $n_j$ and $n_j+d_j$ (with $d_j$ positive integers for all $j$).
	Second, we consider standard (i.e., locally coupled) atoms coupled to the $(1,2)$-Fibonacci waveguide, so that the interaction  reads
	\begin{eqnarray}\label{eq:HintSmall}
		\hat H_{\rm int}
		=
		g
		\sum_{j=1}^{N_e}
		\left(
		\hat \sigma_j
		\hat a_{n_j} ^\dagger
		+
		{\rm H.c.}
		\right)
		.
	\end{eqnarray}
	
	Before describing the features specific to each case, we highlight those which do not depend on the details.
	
	\begin{itemize}
		\item In what follows, we study the resonant case $\Delta=0$, so that giant atoms are resonant with the middle of the multifractal spectrum of $\hat H_{(1,1)}$, and local atoms lie at the middle of the central bandgap of $\hat H_{(1,2)}$.
		
		\item In standard waveguide QED literature \cite{roy2017colloquium, gu2017microwave, sheremet2023waveguide} , decoherence-free interactions are derived by calculating the master equation describing the reduced emitters' dynamics \cite{leonforte2024quantum}. 
		Briefly, tracing out the waveguide degrees of freedom and assuming that decoherence-free conditions hold (in particular for us, either off-resonant standard atoms, or resonant giant atoms with judiciously arranged coupling points), the reduced emitters' state $\varrho$ evolves as \cite{leonforte2024quantum}
        \begin{equation}\label{eq:general_atomic_master_equation2}
			\dot{\varrho}=-i[\hat H_e+ \hat H_{\text{eff}}, \varrho].
		\end{equation}
		with
		\begin{equation}\label{eq:H_effect}
			\hat H_{\text{eff}}
			=
			\sum_{ij}
			\mathcal{K}_{ij}
			\hat \sigma_{i}^\dagger \hat  \sigma_{j},
		\end{equation}
		The waveguide-mediated couplings $\mathcal{K}_{ij}$ between atoms $i$ and $j$ can be written in terms of the waveguide Green's function \cite{economou2006green}. 
		These mediated interactions can be equivalently described in terms of atom-photon bound states \cite{leonforte2024quantum}.
		In this picture, emitter $i$ and $j$ cross-talk thanks to the spatial overlap between the emitter $i$th's position and the virtual photonic cloud (the photonic component of the atom-photon bound state $\ket{\Psi_j}$, see next point) induced by  emitter $j$ (and vice versa), namely
		\begin{eqnarray}\label{eq:Kij}
			\mathcal{K}_{ij}=g\,\langle n_i\ket{\Psi_j}\,.
		\end{eqnarray}
		Although these two pictures (Green's functions and atom-photon bound states) are equivalent routes to deriving the effective coherent interactions in standard waveguide QED, 
		this is not the case with aperiodic waveguides. Indeed, the derivation of the master equation for the reduced emitters' dynamics in a $(p,q)$-Fibonacci waveguide is not analytically tractable \cite{jianxin1993green}. 
		These waveguide Hamiltonians cannot be diagonalized analytically, and their corresponding DOS is nowhere smooth~\cite{jagannathan2021fibonacci}, potentially affecting the validity of the Markov approximation.
		Still, in the resonant case $\Delta=0$, we derive an analytical expression of the
		atom-photon bound states. 
		These states mediate interactions in exactly the same way as in standard waveguide QED, see Eq.~\eqref{eq:Kij}. We support this statement numerically by comparing exact and effective atomic dynamics, cf. Appendix~\ref{app:exactVSnum}.
		In turn, our Fibonacci waveguide QED setups showcase an example, not previously noticed to the best of our knowledge, for which decoherence-free interactions cannot be derived through a master equation, as the waveguide DOS is 
        not smooth (thus breaking Markov approximation).
        The only available route being thus the atom-photon bound states picture.

		\item In both cases we consider $(p,q)=(1,1)$ and $(p,q)=(1,2)$. The atom-photon bound states $\ket{\Psi}$ of our interest fall into the class of VDSs. These are states of the form 
		\begin{eqnarray}\label{eq:vds}
			\ket{\Psi}\propto \varepsilon\ket{e}\ket{\rm{vac}} + \ket{g}\ket{\psi}\,,
		\end{eqnarray}
		with $\ket{\psi}=\sum_n \psi_n \hat a_n^\dagger\ket{\rm{vac}}$, and $\ket{\rm{vac}}$ is the waveguide state without photons, satisfying ${\hat H\ket{\Psi} = \Delta \ket{\Psi}}$. We therefore develop our discussion using the language of VDSs \cite{leonforte2021vacancy}.
        In Eq.~\eqref{eq:vds} and in the remainder of the manuscript we will use the shorthand notation $\ket{e}\equiv \ket{e}\ket{\rm{vac}}$ and $\ket{\psi}\equiv \ket{g}\ket{\psi}$ ($\ket{\psi}$ being a photonic state).
		
	\end{itemize}

    \begin{figure*}
		\centering
		\includegraphics[width=\textwidth]{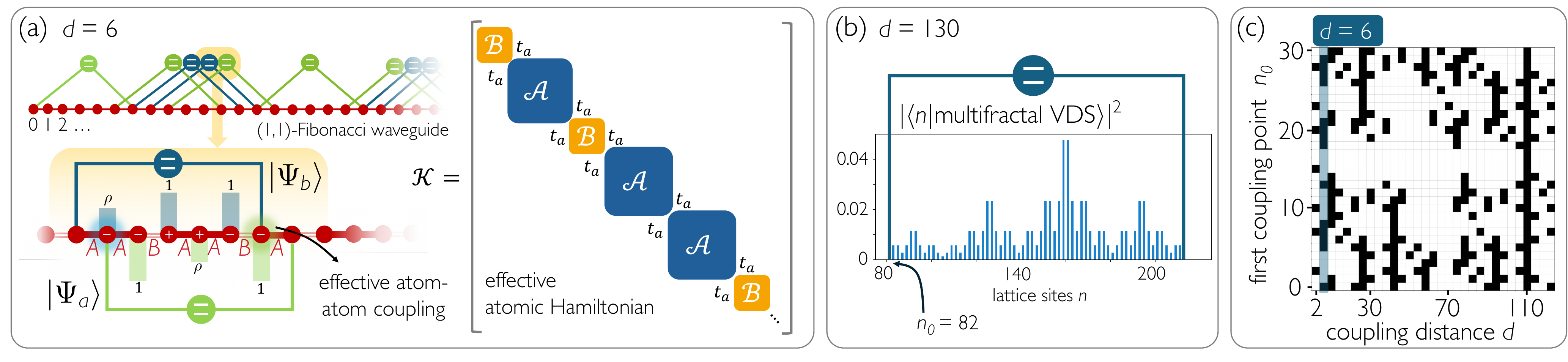}
		\caption{ \textit{Giant emitters in a $(1,1)$-Fibonacci waveguide}. 
        (a) Giant emitters arranged in the first nontrivial configuration  yielding decoherence-free interactions (coupling points at distance $d=6$).
        In this configuration, giant atoms can induce three kinds of atom-photon bound states $\ket{\Psi_{a}}$, $\ket{\Psi_{b}}$ and its mirror symmetric one, whose mutual overlap gives the coherent emitter interactions. On the right we show the block structure of the effective Hamiltonian $\mathcal{K}$ when the giant atoms are coupled to all allowed positions in the $(1,1)$-Fibonacci waveguide, reflecting the underlying Fibonacci aperiodicity.
        (b) Multifractal vacancy-like dressed state induced by a giant atom. When the distance between coupling points is large enough, the multifractality of the photonic component becomes visible.
        (c) Allowed first coupling point position $n_0$ (in black) as a function of distance between coupling points $d$. We only show the distances $d=2,\,6,\,10\,...$ as all other values of $d$ do not yield a VDS.}
		\label{fig:giants}
	\end{figure*}
    
	\subsection{Giant emitters in a $(1,1)$-Fibonacci waveguide}\label{sec:giantsFibwg}
	In the limit of a uniform waveguide ($t_n\equiv t_A$), a single giant atom with coupling points at distance ${d_j=2\,(2\nu+1)}$, with $\nu$ a non-negative integer, induces a bound VDS, which is a bound state in the continuum (BIC) \cite{hsu2016bound, calajo2019exciting}. Note that (i) the distance $d_j$ is independent of $j$, meaning the uniform waveguide's translational invariance is restored, and (ii) the number of sites between the coupling points (excluded) is odd. Such a dressed state features an atomic component and a  photonic excitation trapped in the part of the waveguide array between the coupling points. When many giant atoms are coupled to a uniform waveguide, it is then possible to engineer interactions protected from decoherence, mediated by the BICs seeded by each atom~\cite{leonforte2024quantum}.
	
	We study here the fate of this physics in an aperiodic waveguide. We consider the total Hamiltonian $\hat H$ as in Eq.~\eqref{eq:FibWQEDHam} with the hopping sequence $\{ t_n \}$ obtained from the $(1,1)$-Lucas substitution rule. The interaction Hamiltonian is the one in Eq.~\eqref{eq:HintGiant} (giant-atom case).
	
	Consider first a single on-resonance ($\Delta=0$) giant atom coupled to the resonators at positions $n_0$ and $n_0+d$. Following Ref.~\cite{leonforte2024quantum}, we define the site state
	$\ket{\chi}=\left(\ket{n_0} + \ket{n_0+d}\right)/\sqrt{2}$, with $\ket{n}=\hat a_n ^\dagger \ket{\rm{vac}}$. In this picture the atom is coupled to a single effective site $\chi$. The waveguide Hamiltonian with a vacancy at (i.e, removing) the effective site $\chi$ is labeled as $\hat H_{(1,1),\chi}$.
	For a VDS $\ket{\Psi}$ to exist one  requires that it couples to $\ket{\chi}$ via $\hat H_{(1,1)}$ and decouples from 
	the orthogonal state $\ket{\chi^{\perp}}=\left(\ket{n_0} - \ket{n_0+d}\right)/\sqrt{2}$.
	A zero-energy ($\Delta=0$) eigenstate  $\ket{\psi}$ of $\hat H_{(1,1),\chi}$ (a consequence of $\ket{\Psi}$ being a VDS, see Eq.~\eqref{eq:vds}) can in general only arise if $d=2m$ with positive integer $m$. That is if the number of sites between the coupling points (excluding them) is odd. In this case, it is written (up to normalization) as
\begin{equation}\label{eq:general_psi_on_B_xi}
		\ket{\psi}=
		\sum_{n=0}^{(d-2)/2} (-1)^{n+1}\psi_n\ket{n_0+2n+1},
	\end{equation}
	with coefficients 
    %$\psi_{0}=1$ and 	
    \begin{equation}\label{eq:psi_VDS_giant_coefficient}
		\psi_{n>0} = 
		\frac{t_{n_0+1}}{t_{n_0+2}}
		\cdot
		\frac{t_{n_0+3}}{t_{n_0+4}}
		\cdot ... \cdot
		\frac{t_{n_0+2n-1}}{t_{n_0+2n}}
        \,\psi_{0}.
	\end{equation}
	Thus, the aforementioned conditions for a VDS, namely $\bra{\chi}\hat H_{(1,1)}\ket{\psi}\neq0$ and $\bra{\chi^{\perp}}\hat H_{(1,1)}\ket{\psi}=0$, imply 
	that \begin{eqnarray}\label{eq:conditionGiant}
		\psi_0t_{n_0}=(-1)^{(d-2)/2}\,\psi_{(d-2)/2}t_{n_0+d-1},
	\end{eqnarray}
	which is satisfied only for odd $m$. The allowed distances between coupling points for a bound state to exist in a $(1,1)$-Fibonacci waveguide are therefore the same as those for a uniform waveguide.
	However, lacking translational invariance, for each fixed distance $d$ a dressed bound state can be seeded only for certain specific coupling positions $n_0$, cf.~Fig.~\ref{fig:giants}(c). We now analyze these bound states and the effective interactions which they mediate in a few relevant cases.
	
	In the minimal case $d=2$, Eq.~\eqref{eq:conditionGiant} reduces to ${t_{n_0}=t_{n_0+1}}$. Thus, from all the distinct 2-letters subwords of the $(1,1)$-Fibonacci sequence generated from~\eqref{eq:rule}, only $AA$ allows for a VDS to form as the $BB$ subword does not occur in the  $(1,1)$-Fibonacci sequence. A dressed bound state is then seeded only when the first coupling site is followed by two $t_A$ hoppings and reads $\ket{\Psi}\propto \ket{e} -g/t_A \ket{n_0+1}$.
	As the subword $AAA$ never appears in the $(1,1)$-Fibonacci sequence, the bound states induced by any pair of giant atoms coupled to the waveguide cannot overlap. Hence, decoherence-free interactions between giant atoms are absent for $d=2$.
	
	The first nontrivial case is $d=6$, see Fig.~\ref{fig:giants}. In this case, the lattice between the coupling points (excluded) is made of ${d-1=5}$ sites, corresponding to a sequence of ${d-2=4}$ hoppings. For the $(1,1)$-Fibonacci waveguide, there are five unique sequences of length four \cite{perrin2012note}. These are
	$BAAB$,
	$ABAA$, 
	$ABAB$, and the mirror symmetric ones ($AABA$, $BABA$).
	The hoppings appearing in Eq.~\eqref{eq:conditionGiant}, $t_{n_0}$ and $t_{n_0+d-1}$ (namely, those connecting to the coupling-point positions), correspond to additional letters at the beginning and the end of these subwords 
    and they must fulfill the constraint $\psi_0 t_{n_0} = \psi_{2}t_{n_0+5}$. Thus, the available candidate subwords for dressed bound states are
	(a) $\cdot A|BAAB|A\cdot $,
	(b) $\cdot A|ABAA|B\cdot $,
	(c) $\cdot B|ABAA|B\cdot $,
	(d) $\cdot A|ABAB|A\cdot $,
	and the mirror symmetric ones (the central dots represent the coupling points positions while the vertical bars are meant to guide the eye).
	These are constructed from the five aforementioned words, considering that the subwords $BB$, $AAA$, and $BABAB$ never appear in the $(1,1)$-Fibonacci sequence \cite{mignosi2002words}.
	The case $t_{n_0}=t_{n_0+5}$ corresponds to the subword (a) $\cdot A|BAAB|A \cdot$. If instead $t_{n_0}\neq t_{n_0+5}$, the only two other remaining subwords are (b) $\cdot A|ABAA|B \cdot$ and its mirror symmetric one. 
    The words (c) and (d) do not satisfy the constraint.
	In total, there are then two different subwords (up to mirror symmetry) yielding a VDS (still for $d=6$ and $\Delta=0$). These mutually overlap, such that $d=6$ is the minimal example for which giant atoms coupled to the $(1,1)$-Fibonacci waveguide can coherently interact.
    The corresponding VDSs are 
    \begin{equation}
    \medmath{
        \ket{\Psi_a}
        =
        \frac{1}{\mathcal N}
        \left[
        \ket{e}
        -
        \frac{g}{t_A}
        \left(
        \ket{n_0+1}
        -\rho \ket{n_0+3}
        +\ket{n_0+5}
        \right)
        \right]
        \nonumber
        }
    \end{equation}
    occurring for instance for $n_0=5$, and 
    \begin{equation}
    \medmath{
        \ket{\Psi_b}
        =
        \frac{1}{\mathcal N}
        \left[
        \ket{e}
        -
        \frac{g}{t_B}
        \left(
        \rho\ket{n_0+1}
        -\ket{n_0+3}
        +\ket{n_0+5}
        \right)
        \right]
        \nonumber
        }
    \end{equation}
    occurring for instance for $n_0=7$ (note that $n_0=5,\,7$ are two allowed coupling positions for $d=6$, cf.~Fig.~\ref{fig:giants}(c)). For both normalization constants we have 
	$\mathcal N \sim 1 + \mathcal{O}(g^2)$.
	Remarkably, the effective giant emitters' Hamiltonian, Eq.~\eqref{eq:H_effect}, inherits the aperiodicity of the underlying waveguide. Indeed, as shown in Fig.~\ref{fig:giants}(a), the matrix ${\mathcal K = \{ \mathcal K_{ij} \}}$ is made up by the blocks $\mathcal{B}=\{0\}$ 
	and 
	\begin{equation}\label{eq:A_Fibonacci}
		\mathcal{A}=
		\begin{pmatrix}
			0 & t_a & 0 & t_c\\
			t_a & 0 & t_b & 0\\
			0 & t_b & 0 & t_a\\
			t_c & 0 & t_a & 0
		\end{pmatrix},
	\end{equation}
    with $t_a=g^2/t_A$, $t_b=g^2/t_B$, and $t_c=-\rho\,t_a$,
	coupled by the hopping rate $t_a$. These blocks alternate as in the Fibonacci sequence, see Fig.~\ref{fig:giants}(a) right column.
	
	The vacancy-like bound states induced by giant atoms in the $(1,1)$-Fibonacci waveguide are a generalization of the BICs in standard waveguide QED. As the waveguide spectrum is not a continuum, we have avoided referring to them as such. It would be tempting to name them \textit{multifractal bound states}, however, multifractality refers to the scaling properties of a state, implying that the support of the state should be reasonably large. 
	It is meaningless to attribute multifractal features to the states we discussed for $d=2$ and $6$. Nevertheless, coupling a giant atom at two distant points ($d\gg 1$) to the waveguide and considering that the photonic part of a VDS is nothing but the zero-energy state of the waveguide with a vacancy~\cite{leonforte2021vacancy} (which is multifractal itself), we can call the induced vacancy-like state a \textit{multifractal VDS} (the term `bound' seems inappropriate given the large size of the state support). We show an example in Fig.~\ref{fig:giants}(b).
	
    Finally, we note that, applying the condition of Eq.~\eqref{eq:conditionGiant}, for each given distance $d$ between the coupling points, we directly obtain the first coupling point $n_0$ which yields a VDS (for $\Delta=0$). The resulting map is shown in Fig.~\ref{fig:giants}(c).
	
	\subsection{Local emitters in a $(1,2)$-Fibonacci waveguide}
	We now consider locally coupled emitters, cf.~Eq.~\eqref{eq:HintSmall}, to a $(1,2)$-Fibonacci waveguide. We assume that the emitters are off-resonant and tuned to the middle of the bandgap ($\Delta=0$).
	As we mentioned previously, this setup can be seen as the aperiodic version of the SSH waveguide QED setup in \cite{bello2019unconventional}, as we outline below.
	
	In the previous section about giant atoms in a $(1,1)$-Fibonacci waveguide, the length of the waveguide never played a role. Indeed, the bound VDSs in that case have support only between the coupling points of the giant atom, making the remainder of the waveguide irrelevant.
    In contrast, since the bound VDS induced by a locally coupled atom is not strictly confined, we must first address the boundary conditions of the waveguide.
	
	As we mentioned in Section~\ref{sec:fibwg}, Fibonacci waveguides are always effectively  under open boundary conditions. Therefore, we make a few assumptions to render the $(1,2)$-Fibonacci waveguide as clean as possible. Namely, we aim to prevent the emitters from being resonant with unwanted zero-energy states (to avoid hybridization with edge modes \cite{bello2019unconventional}). More simply, we require the $(1,2)$-Fibonacci waveguide to be gapped without mid-gap zero-energy states.
	To this end, we assume that the number of resonators is even (otherwise a zero-energy state always exists because of chiral symmetry), and that $t_0=t_A>t_B$, which guarantees the absence of (almost-)zero-energy edge states. 
	Last, we assume that the number of resonators is large to the extent that atoms can be coupled to the bulk of the waveguide.
	
	Coupling the $j$th atom to the resonator $n_j$, a zero energy atom-photon bound state emerges. The explicit form depends on the parity of $n_j$.
	If $n_j$ is odd the state reads
	\begin{eqnarray}
    \medmath{
		\ket{\Psi_j^{\rm odd}}
		=
		\frac{1}{\mathcal N}
		\left[
		\ket{e}
		+
		\frac{g}{t_A}
		\sum_{n=0}^{\frac{n_j-1}{2}}
		(-1)^{n+1}
		\rho^{\beta_j(n)}
		\ket{n_j-1-2n}
		\right]}
		\nonumber
	\end{eqnarray}
	with $\beta_j(n)\equiv \beta(n_j-1-2n,n_j-1)$ and $\beta(n,m)$ is the number of $t_B$-hoppings between site $n$ and $m$. 
	Notice that the photonic component has  non-zero amplitudes only on even sites and to the left of the atom.
	If instead  $n_j$ is even we get
	\begin{eqnarray}
    \medmath{
		\ket{\Psi_j^{\rm even}}
		=
		\frac{1}{\mathcal N}
		\left[
		\ket{e}
		+
		\frac{g}{t_A}\!\!\!
		\sum_{n=0}^{ \frac{N-2-n_j}{2}}
		\!\!\!\!(-1)^{n+1}
		\rho^{\beta'_j(n)}
		\ket{n_j+1+2n}
		\right]}
		\nonumber
	\end{eqnarray}
	with $\beta'_j(n)\equiv \beta(n_j+1,n_j+1+2n)$.
	In this case, the photonic component has  non-zero amplitudes only on odd sites and to the right of the atom.
	For both normalization constants we have 
	$\mathcal N \sim 1 + \mathcal{O}(g^2)$.
	
\begin{figure}
		\centering
		\includegraphics[width=\columnwidth]{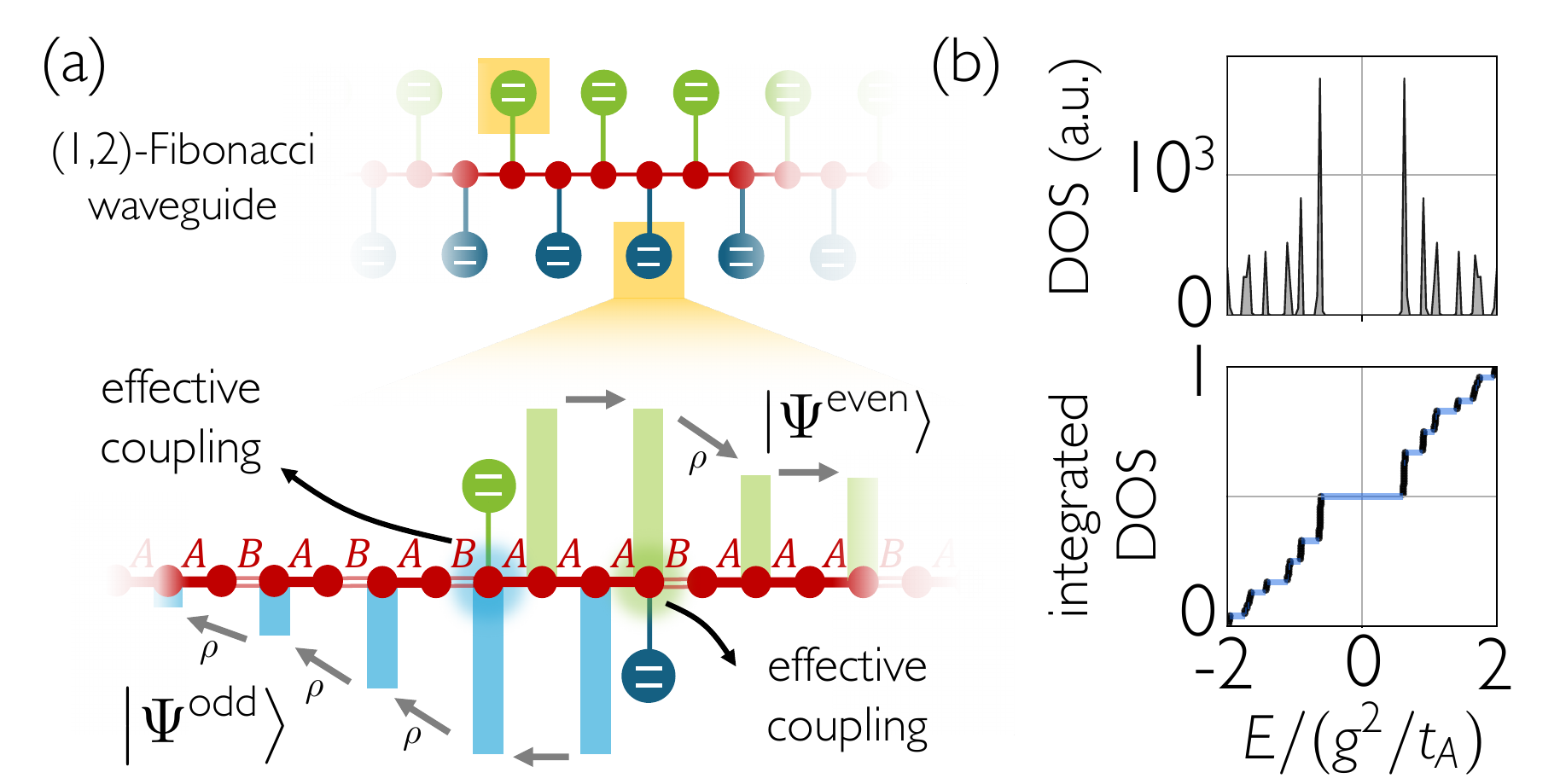}
		\caption{ \textit{Emitters in a $(1,2)$-Fibonacci waveguide}. 
        (a) Local off-resonant emitters yield chiral atom-photon bound states whose photonic profile decays exponentially with aperiodic modulation.
        (b) DOS and integrated DOS of the effective emitter Hamiltonian showing its critical behavior. Parameter values: $t_B=0.2\,t_A$, $g=0.05\,t_A$.
        }
		\label{fig:small}
	\end{figure}

	The profile of these bound states (namely the photonic component) reveals the connection between the $(1,2)$-Fibonacci waveguide and the SSH model \cite{bello2019unconventional}: the dressed states $\ket{\Psi_j^{\rm even/odd}}$
	include the SSH limit for which ${\beta_j(n)=\beta'_j(n)=n}$, and the uniform waveguide limit with $\beta_j(n)=\beta'_j(n)=0$.
	Remarkably, the photonic part of these dressed states inherits the aperiodicity of the underlying waveguide: Assuming that $\psi_0$ is a non-zero photonic component, the following non-zero photonic component will be (i) $\rho\psi_0$ if a $t_B$ coupling is met on the way, or (ii) $\psi_0$ if two $t_A$ couplings are met, see Fig.~\ref{fig:small}(a).
    We note that this is true for any gapped $(p,q)$-Fibonacci waveguide, cf.~Fig.~\ref{fig:gapsize}, and it is not restricted to the specific case $(p,q)=(1,2)$.
	Finally, the aperiodic structure of the dressed states translates itself into the multifractality of the effective atom-atom Hamiltonian $\mathcal{K}_{ij}=\langle n_i\ket{\Psi_j^{\rm even-odd}}$, which can be seen from its DOS and integrated DOS in Fig.~\ref{fig:small}(b).

    \section{Conclusion}\label{sec:conc}

    In this work, we have proposed and analyzed a platform for waveguide QED utilizing one-dimensional photonic arrays with quasiperiodic hopping strengths, governed by the Fibonacci-Lucas substitution rule. We have shown that these ``Fibonacci(-Lucas) waveguides'' provide a unique environment for light-matter interactions, characterized by multifractal energy spectra and critical eigenstates. These stand in stark contrast to conventional periodic systems.

    Our central result demonstrates the ability to engineer complex, decoherence-free interactions in these aperiodic structures. We investigated two paradigmatic systems. For giant atoms coupled to a $(1,1)$-Fibonacci waveguide, we found that the formation of mediating atom-photon bound states is exquisitely sensitive to the local aperiodic structure of the hoppings and therefore reflecting statistical properties of subwords of the aperiodic sequence. This leads to an effective atomic Hamiltonian whose interaction matrix itself follows a Fibonacci sequence, imprinting the waveguide's complexity onto the emitters' dynamics. For local atoms in the gap of a $(1,2)$-Fibonacci waveguide, we showed that the mediating bound states have aperiodically modulated tails, giving rise to an effective Hamiltonian with long-range interactions and multifractal properties. A crucial aspect of our analysis is the use of the bound-state approach for deriving coherent interactions in a system in which the singular density of states precludes a standard master equation formalism.

    Our findings open several avenues for future research. The rich structure of Fibonacci-Lucas sequences, controlled by integers $(p,q)$, offers a vast, tunable family of quasiperiodic environments, allowing for the design of a wide range of interaction patterns beyond those studied here. An immediate next step is to explore the \textit{dissipative} dynamics in these systems: How do collective phenomena, such as collective and correlated radiance \cite{li2025supercorrelated}, manifest in a multifractal environment? 
    Recently, 
    quantum state transfer between spins coupled to the opposite sides of a Fibonacci chain has been studied~\cite{ghosh2025quantum}. 
    The extension to Fibonacci-Lucas waveguides and whether they can be used for quantum pumps~\cite{ghosh2025quantum, citro2023thouless} is an open question.
    Furthermore, quasiperiodic arrays, including the Fibonacci waveguides studied here, may host unique topological properties~\cite{kraus2012topological}. The possibility of defining topological invariants~\cite{LinPRB2021} and how  topological properties might impact the mediated interactions between emitters remains an open  question. 
    From a practical standpoint, the proposed setups are within reach of current experimental platforms, particularly superconducting circuits and coupled resonator arrays in which hopping amplitudes can be precisely engineered~\cite{ferreira2021collapse, jouanny2025high, kim2021quantum, zhang2023superconducting, scigliuzzo2022controlling, owens2022chiral, carusotto2020photonic}. The ability to create long-range interactions with quasiperiodic ``plateaux'' 
    offers a new tool for quantum simulation and the creation of robust, topologically-inspired quantum states. By bridging the gap between perfectly ordered and fully disordered systems, Fibonacci waveguide QED provides a new avenue for controlling light and matter in complex quantum environments.

\section*{Acknowledgments}
	We thank Francesco Ciccarello for helpful comments on the manuscript. We acknowledge funding from the Max Planck
	Society's Lise Meitner Excellence Program 2.0.
	F.R.~acknowledges financial support by the European Union-Next Generation EU with the project ``Quantum Optics in Many-Body photonic Environments'' (QOMBE) code SOE2024\_0000084-CUP B77G24000480006.

\bibliographystyle{quantum}
% \bibliography{references}

\onecolumn
\appendix

	%Appendix A
	\section{Relation between the SSH, (1,1)- and (1,2)-Fibonacci waveguides}\label{app:SSH1112}
\begin{figure}
		\centering
		\includegraphics[width=0.75\columnwidth]{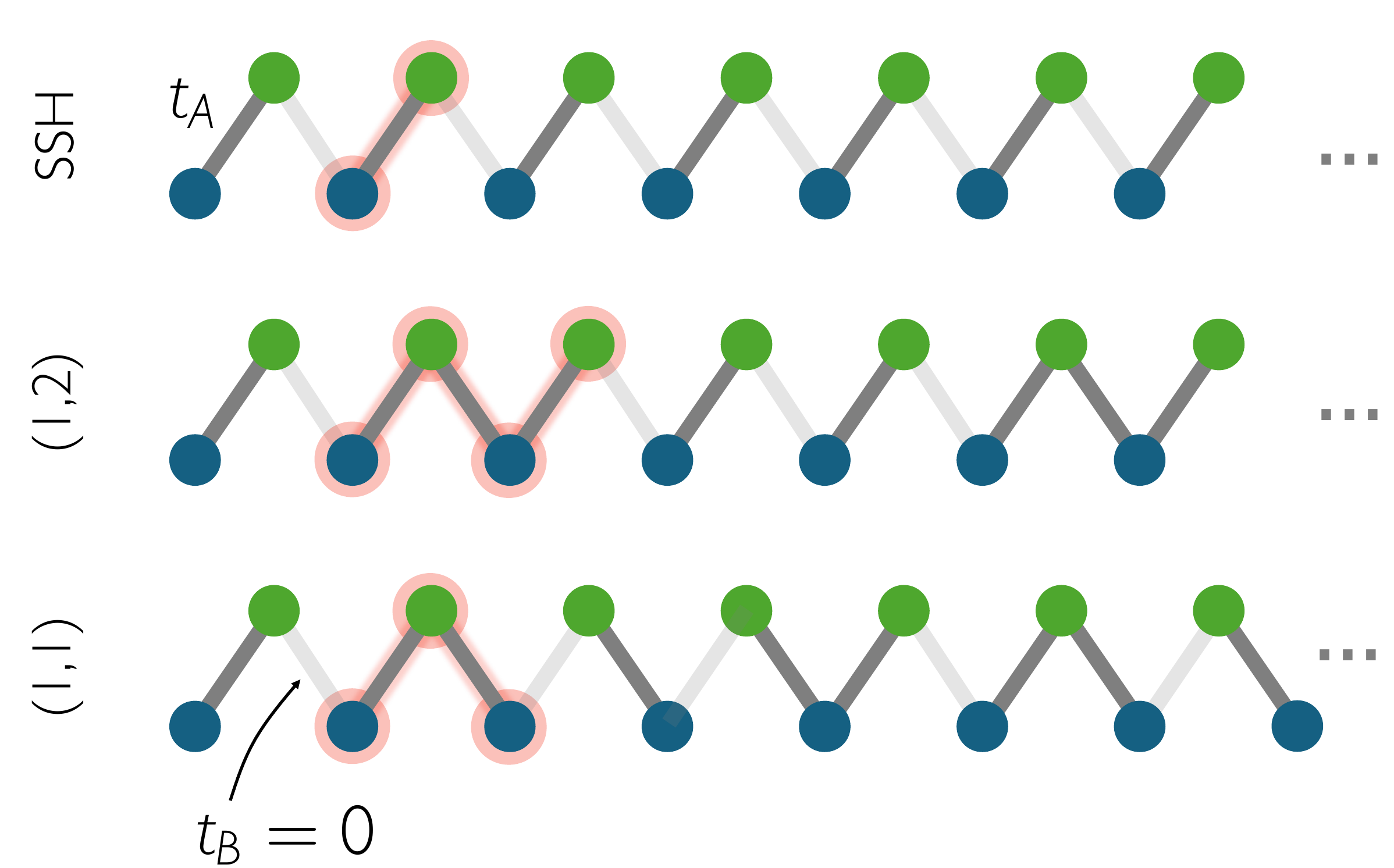}
		\caption{\textit{SSH vs Fibonacci waveguides in the uncoupled limit}. 
				The SSH waveguide (top row), the $(1,2)$-Fibonacci waveguide (middle row), and the $(1,1)$-Fibonacci waveguide (bottom row) in the $t_B=0$  limit.
				$(p,q)$-Fibonacci waveguide with odd $p$ and even $q$ have a topological gap around zero energy. In the limit of $t_B=0$, these chains consist of groups of sites with even number (red), such that they only posses degenerate energy bands at $\pm t_A$. A weak perturbation of non-zero $t_B$ smoothly modifies these bands without closing the gap. In contrast, all other Lucas chains, for example the $(1,1)$-Lucas chain, have groups of even and odd number of sites (red) in the dimerized limit, hence, they posses also a degenerate energy band at $E=0$, so that no gap around zero energy can exist.
		}
		\label{fig:topology}
	\end{figure}

	In order to clarify the relation between
	the SSH, the $(1,2)$-Fibonacci, and the $(1,1)$-Fibonacci waveguides we consider the limit $t_B=0$, see Fig.~\ref{fig:topology}. In this limit, both the SSH and the $(1,2)$-Fibonacci waveguide solely consist of $n$-mers made from an even number of sites (dimers, tetramers,...). These waveguides have degenerate energy bands at $E=\pm t_A$. A small perturbation from a non-zero $t_B$ will smoothly modify these bands without closing the gap at $E=0$. In contrast, the $(1,1)$-Fibonacci chain consists of $2n$-mers as well as trimers. Hence, this waveguide has two degenerate bands at $E=\pm t_A$ and one at $E=0$. Therefore, especially with a non-zero $t_B$, no gap around zero energy can exist. This argument holds for all $(p,q)$-Fibonacci waveguides with both $p$ and $q$ even (or both odd).

	%Appendix B
	\section{On multifractality}\label{app:cantor}
\begin{figure}
		\centering
		\includegraphics[width=0.75\columnwidth]{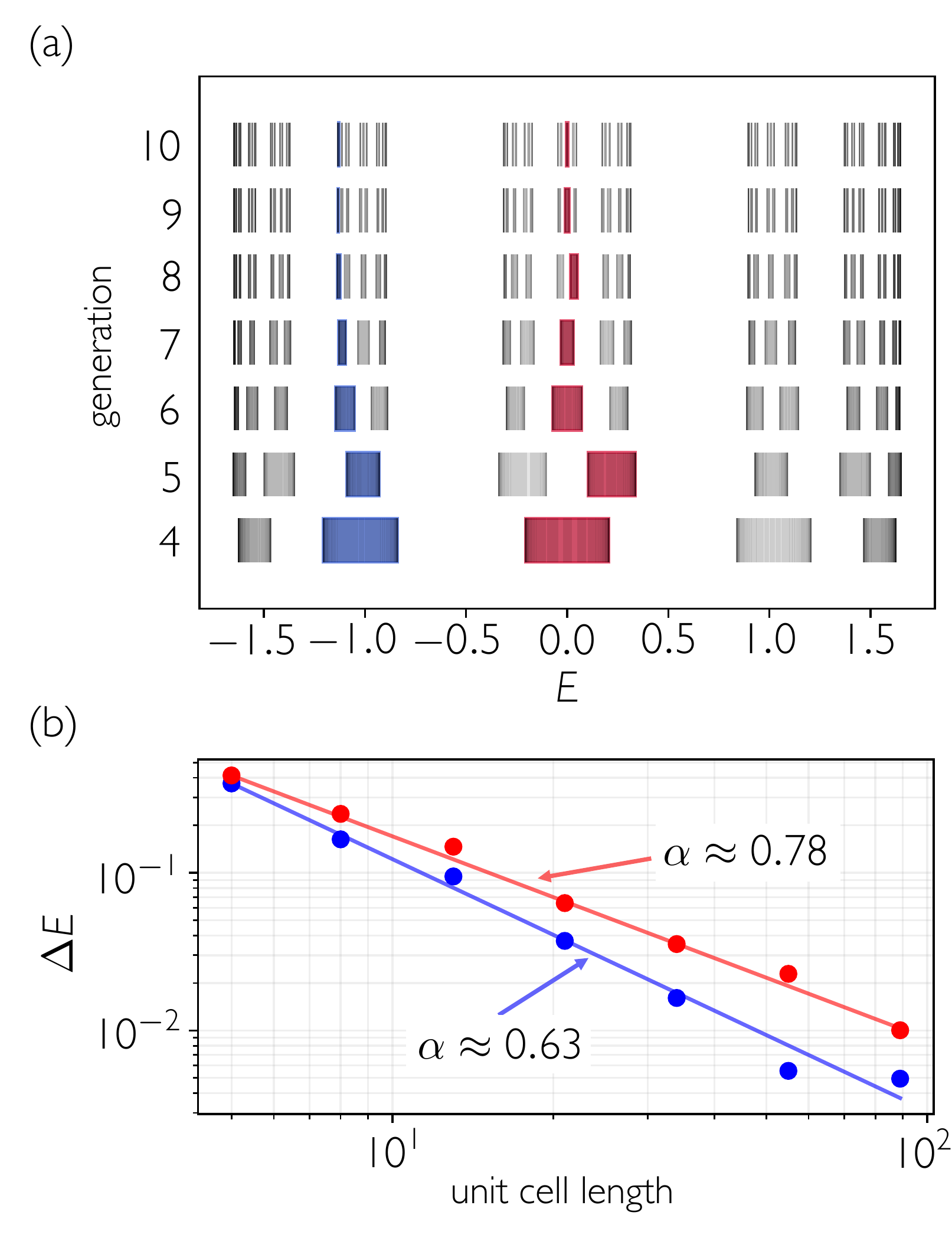}
		\caption{\textit{Multifractal scaling of periodic Fibonacci chain approximants}. Energy spectrum for generations $n=4$ to $n=10$ with hopping ratio $\rho = t_B/t_A = 0.5$. (a) Energy spectrum evolution showing the hierarchical splitting of bands. Two distinct  branches are tracked: the central band (shaded red at an integrated DOS $\mathcal{N}(E)=0.5$) and a branch below the main gap (blue shaded at an integrated DOS $\mathcal{N}(E) \approx 0.25$). (b) Finite-size scaling of the bandwidth $\Delta E$ versus the unit cell length $F_n$ on a log-log scale. The solid lines represent power-law fits $\Delta E \sim L_n^{-1/\alpha}$. The distinct scaling exponents display the multifractal nature of the energy spectrum.}
		\label{fig:cantor}
	\end{figure}

	While the structure of the integrated DOS, presented in Fig.~\ref{fig:dos} suggests a multifractal spectrum, a global view alone cannot distinguish between a simple monofractal (for which the scaling dimension is constant) and a multifractal (for which the dimension varies locally) behavior. To address this, we track the splitting of specific spectral bands and track their band widths $\Delta E$ as paths through the generations in Fig.~\ref{fig:cantor}(a), where for each generation, we used an overall system of approximately $2000$ sites and approximants of unit cell size $F_n$ ($F_n$ being the $n$th Fibonacci number). In particular, we focus on the path of the center band ($\mathcal{N}(E)=0.5$, red) and a second region below the primary gap ($\mathcal{N}(E) \approx 0.25$, blue). We characterize these paths of splitting bands using the scaling $\Delta E \sim F_n^{-1/\alpha}$. If the spectrum were a regular fractal, the scaling of the bandwidths $\Delta E$ with system size $F_n$ would be identical for both paths and Fig.~\ref{fig:cantor}(a) would be a normal Cantor set. However, the log-log scaling analysis in Fig.~\ref{fig:cantor}(b) reveals a clear difference in their behavior. The central band scales with an exponent $\alpha \approx 0.78$. In contrast, the other path scales with a larger exponent $\alpha \approx 0.63$. This measurable variation in the local scaling exponent $\alpha$ confirms that the Fibonacci spectrum is indeed multifractal, characterized by a distribution of singularity strengths rather than a single global fractal dimension.

	%Appendix C
	\section{Central gap of $(p,q)$-Fibonacci waveguides}\label{app:centralgap}
	
\begin{figure}[h]
		\centering
		\includegraphics[width=0.5\columnwidth]{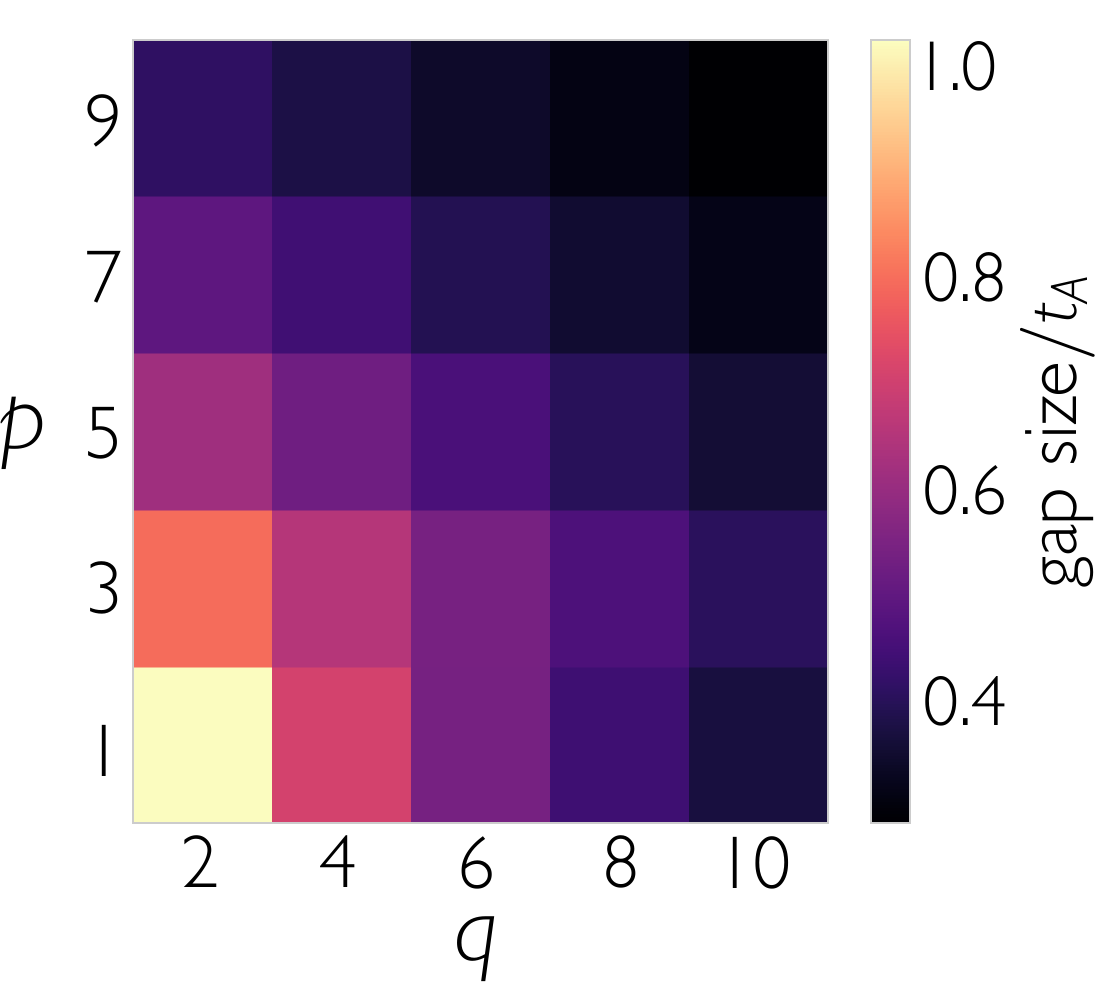}
		\caption{ \textit{Gap size}. Size of the central bandgap of $(p,q)$-Fibonacci waveguides as a function of $p$ and $q$ (the values of $p$ and $q$ not displayed correspond to gapless waveguides). We used  $t_B=0.2\,t_A$.}
		\label{fig:gapsize}
	\end{figure}

	All $(p,q)$-Fibonacci waveguides display fractal gaps, but some of them feature a central gap, which does not have a multifractal origin (it generalizes the SSH bandgap, see Fig.~\ref{fig:topology} and Fig.~\ref{fig:dos}).
	We numerically computed the size of the central bandgap as a function of $p$ and $q$, and show it in Fig.~\ref{fig:gapsize}.

	%Appendix D
	\section{Singularity spectrum}\label{app:falpha}
	The singularity spectrum $f(\alpha)$ for the wavefunctions presented in this work was computed numerically using a box-counting method~\cite{chhabra1989direct, thiem2013wavefunctions}.
	For a selected normalized eigenstate $\ket{\psi}$ on a lattice of size $N$, the probability measure at each site is defined as $\mu_j = |\psi_j|^2$, with $\psi_j=\braket{j}{\psi}$. The idea of the method is to analyze how this measure is distributed at different length scales. Our implementation follows these steps:
	\begin{enumerate}
		\item The lattice is partitioned into $N_b = N/l$ non-overlapping boxes of size $l=2,4,8,16,...$.
		\item We calculate the probability of the wavefunction inside box $k$ of size $l$ as ${P_k(l) = \sum_{j \in \text{box } k} \mu_j}$.
		\item A partition function is calculated by summing the moments of these probabilities as
		\begin{equation}
			Z(q, l) = \sum_{k=1}^{N_b} [P_k(l)]^q,
		\end{equation}
		with $-8<q<8$. To circumvent numerical instabilities, we write 
		$\ln\left(Z(q,l))\right)=\ln\left( \sum_{k=1}^{N_b} e^{q\cdot\ln[P_k(l)]}\right)$ and make use of 
		\begin{equation}
			\ln\left(\sum_k e^{x_k} \right)=M+\ln\left(\sum_k e^{x_k-M} \right),
		\end{equation}
		with $M=\max_k(x_k)$.
		\item The scaling exponent $\tau(q)$ is determined from the relationship $Z(q,l) \sim l^{\tau(q)}$. This is achieved by performing a linear regression of $\ln Z(q,l)$ versus $\ln l$.
		
		\item The singularity spectrum $f(\alpha)$ is then obtained via the standard Legendre transform of the numerically calculated $\tau(q)$
		\begin{align}
			\alpha(q) &= \frac{d\tau(q)}{dq} \\
			f(\alpha) &= q \alpha(q) - \tau(q).
		\end{align}
	\end{enumerate}
	
	We note that this method is a numerical approach distinct to the method used in \cite{holzer1991multifractal} to find an exact solution of $f(\alpha)$ for the Fibonacci chain: The analytical approach analyzes the scaling with system size, while our method tracks scaling with box size $l$.
	
	Finally, we note that the eigenstates in the localized AAH phase possess exponentially decaying tails, i.e., $|\psi_j|\sim \exp(-|j-j_0|/\xi)$. For negative moments $q$, the partition function $Z(q,l)$ is dominated by the small probability measures found in these tails. The resulting large values of $\alpha_{\max}$ are therefore not numerical artifacts and a direct mathematical signature of this exponential decay.

    \begin{figure}[h]
		\centering
		\includegraphics[width=0.75\columnwidth]{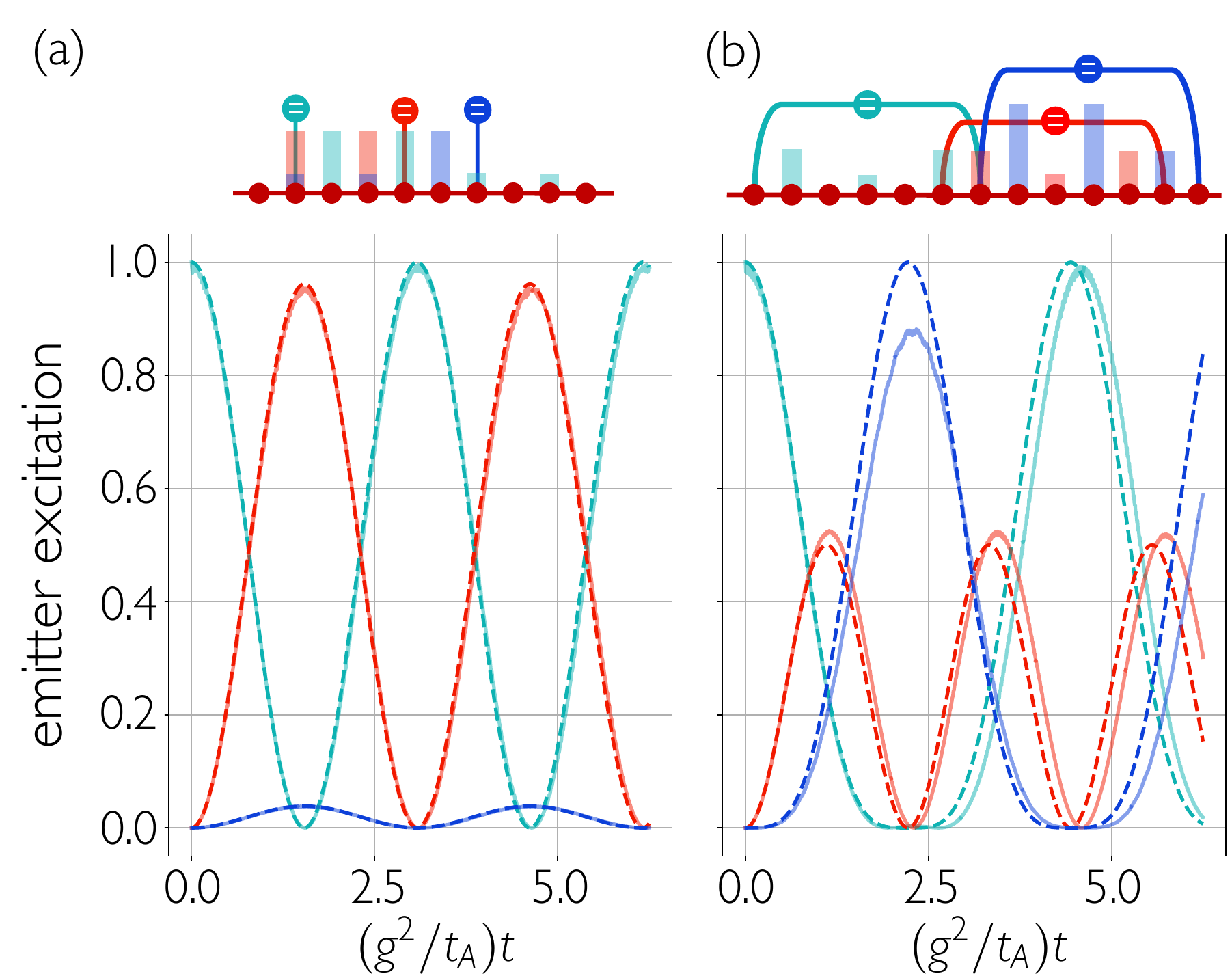}
		\caption{ \textit{Exact vs effective dynamics}.
        Dynamics of the emitters' populations for three emitters starting from the state $\ket{e}_1$ (only the first emitter is excited). Dashed lines correspond to exact dynamics, $|_j\bra{e}e^{-i \hat H t}\ket{e}_j|^2$, while solid lines to effective dynamics, $|_j\bra{e}e^{-i \hat H_\text{eff} t}\ket{e}_j|^2$. First, second, and third emitters correspond to red, turquoise and blue. (a): Three local atoms coupled at $n_0=10$, $n_0=13$, and $n_0=15$ to the $(1,2)$-waveguide with $\Delta=0$, $t_B=0.2\, t_A$, and $g=0.05\,t_A$. There is no effective connection between the second and third atom. (b) Three giant atoms coupled at $n_0=34,\,39,\, 40$ ($d=6$) to the $(1,1)$-waveguide with the same parameter values. Energy in (b) oscillates between the first (turquoise) and third atom (blue) via the second atom (red).}
		\label{fig:numerics_vs_H_eff}
	\end{figure}

    %Appendix E
	\section{Exact vs effective dynamics}\label{app:exactVSnum}
    We numerically compare the dynamics of the atomic populations (starting with an initially excited emitter) considering the full emitters-waveguide Hamiltonian and only the effective emitter Hamiltonian mediated by the atom-photon bound states, see Fig.~\ref{fig:numerics_vs_H_eff}. 
    The effective atomic systems corresponding to the two cases in Fig.~\ref{fig:numerics_vs_H_eff} (a) and (b) are
    \begin{equation}
        \mathcal{K}^{(a)}=
        \begin{pmatrix}
			0 & 0 & t_c\\
			0 & 0 & t_a\\
			t_c & t_a & 0
		\end{pmatrix},
    \end{equation}
    and
    \begin{equation}
        \mathcal{K}^{(b)}=
        \begin{pmatrix}
			0 & t_a & 0\\
			t_a & 0 & t_a\\
			0 & t_a & 0
		\end{pmatrix},
    \end{equation}
    respectively.

\end{document}